\PassOptionsToPackage{draft}{hyperref}

\documentclass[conference,letterpaper]{IEEEtran}
\ifCLASSINFOpdf
\else
\fi
\usepackage{graphicx}
\usepackage{cite}
\usepackage{algorithm}
\usepackage[noend]{algpseudocode}
\usepackage{algorithmicx}
\usepackage{subcaption}
\usepackage{tabularx}
\usepackage{multirow}

\usepackage{amsmath}
\usepackage{amssymb}

\usepackage[table,xcdraw]{xcolor}
\usepackage[normalem]{ulem}
\useunder{\uline}{\ul}{}
\usepackage[utf8]{inputenc}
\usepackage{booktabs}
\usepackage{longtable}
\usepackage{array}
\usepackage[margin=0.5in]{geometry}
\usepackage{supertabular}
\usepackage{amsmath}
\usepackage{graphicx}

\usepackage{url}
\begin{document}
%
\title{Cybersecurity of Electric Vehicle Charging Infrastructure: Recent Advances, Open Challenges, and Future Directions
}
%
%
%

\author{Joshua~Bean and~Dimitrios~Michael~Manias \\ The Department of Computer Science and Engineering, Mississippi State University\\ jab1896@msstate.edu, dmanias@cse.msstate.edu}

%
%

\markboth{IEEE HPSR 2026}%
{Bean \MakeLowercase{\textit{et al.}}:Catchy title goes here}
%



\maketitle

\begin{abstract}
Electric Vehicles (EVs) have emerged as significant disruptors in the transportation sector over the past decade. Their growing popularity and adoption are accompanied by capital expenditures to deploy charging infrastructure. EV charging infrastructure sits at the intersection of the power grid, the network, and the vehicular client, creating an attractive surface for cyberattacks. Many machine learning-based cybersecurity countermeasures have been developed using various public and private datasets. These countermeasures, often intrusion detection systems, are limited in performance by the quality and expressivity of the training data. This work explores the most common datasets and modeling methods, identifies key limitations and open challenges, and proposes future directions to continue catalyzing innovation in the field. By addressing these data limitations, intrusion detection systems are better positioned to address the constantly evolving cyberthreat landscape of EV charging infrastructure.
\end{abstract}

\begin{IEEEkeywords}
Electric Vehicles, Charging Infrastructure, Cybersecurity, Network Security, Smart Grid Security
\end{IEEEkeywords}

\maketitle

\section{Introduction}

\IEEEPARstart{I}{n} recent years, Electric Vehicle (EV) charging ecosystems have seen rapid development, deployment, and growth worldwide. With many jurisdictions offering incentives to promote EV adoption, the growth seen over the past decade is expected to continue and accelerate. This growth raises significant concerns regarding the safety and security of charging infrastructure worldwide. Specifically, the integration of this infrastructure with critical subsystems, including smart grids, payment systems, and vehicular clients, creates a new target for adversaries to infiltrate. To this end, securing this new threat interface is a critical priority and must be considered proactively instead of reactively as the widespread deployment of these systems continues. 

\begin{figure*}[!t]
\centering
    \includegraphics[width=1.7\columnwidth]{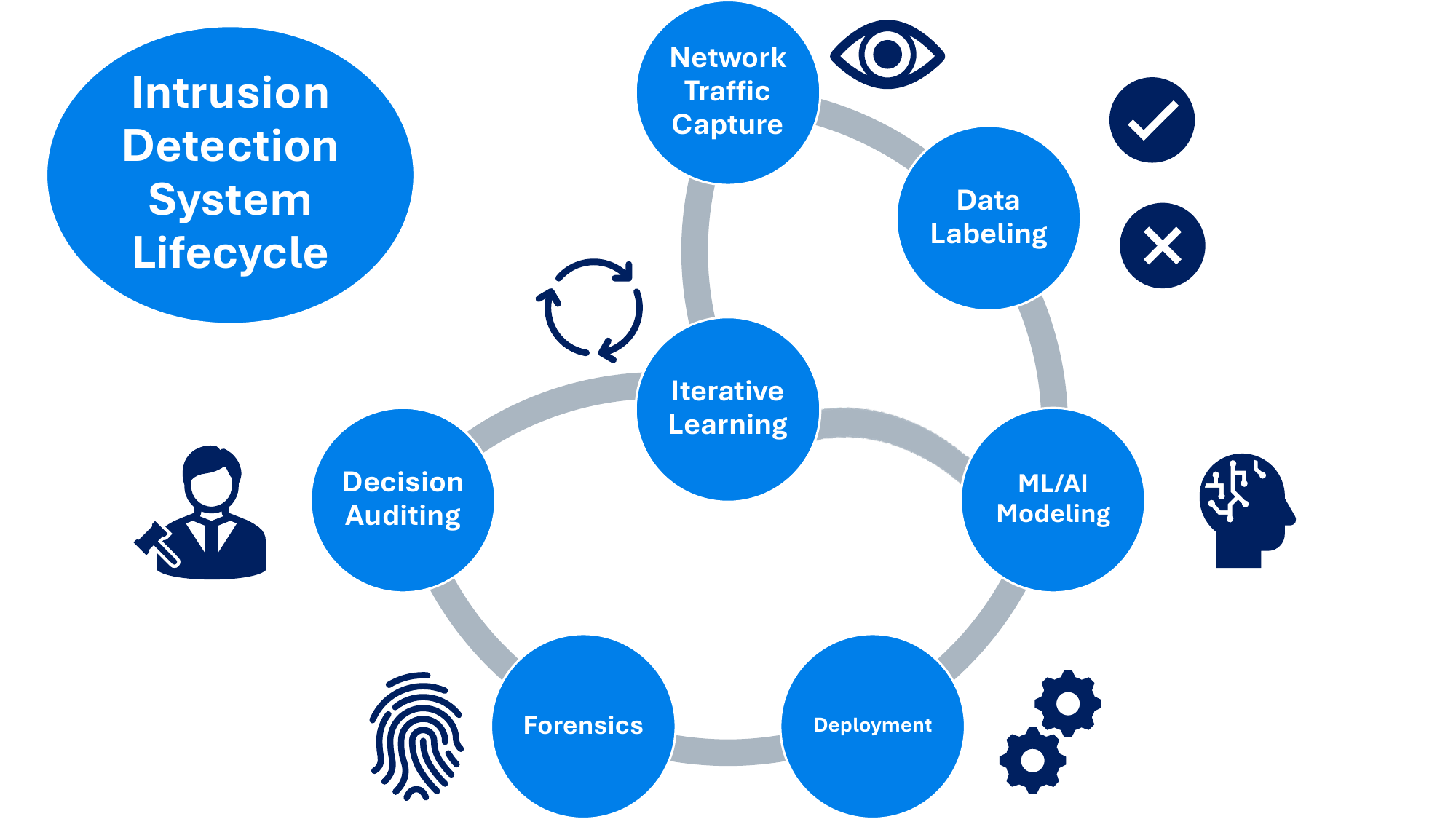}
    \caption{IDS Lifecycle}
    \label{fig:ids-life}
\end{figure*}

Intrusion Detection Systems (IDSs) have emerged as key pillars of EV cybersecurity. These systems are real-time defense mechanisms that protect networked infrastructure from active attacks. However, developing these systems is not a trivial task but rather a multi-step, iterative process. This process begins by capturing network traffic on interfaces and building a corpus of training data for Machine Learning (ML) and Artificial Intelligence (AI) modeling. The network data is then processed and labeled as malicious or benign. Given the volume, velocity, and variability of the data, labeling is incredibly difficult without automation and is prone to mislabeling due to human error (manual) or bounded performance (automatic). ML/AI models are then trained to distinguish between malicious and benign network traffic in real time. The final step in this process is post-incident forensics, which entails IDS decision auditing, threat hunting, and iterative model training. The various components of the IDS lifecycle are depicted in Fig. \ref{fig:ids-life}.

\subsection{Related Work}
Various works have surveyed the state of the art in connected vehicle cybersecurity, with an emphasis on EV systems. These surveys are often broad in scope, providing threat taxonomies or synthesizing detection methods and countermeasure strategies from the literature. Regarding methods and countermeasures, the field is heavily dominated by ML/AI-based implementations, with IDSs, binary and multi-class classification, and anomaly detection frameworks being the most commonly encountered. Given the prevalence of intelligence methods, many existing works do not critically analyze the quality of the datasets used to train the models, particularly with respect to their breadth and depth of coverage, diversity of operating conditions, and representativeness of real-world scenarios. Since data is paramount in ML/AI implementations, understanding the current limitations of existing datasets and the extent to which models trained on them can learn meaningful relationships and achieve real-world deployment feasibility is critical, yet often overlooked. 
\subsection{Contributions}
The contributions of this work can be summarized as follows:
\begin{itemize}
    \item A synthesis of the most prominent datasets used for ML/AI-based cybersecurity within the EV charging infrastructure ecosystem.
    \item A characterization of the types of models and approaches to intrusion detection and anomaly detection found in the literature.   
    \item A critical analysis of the current state of the art, revealing fundamental limitations in dataset realism and operational feasibility, and providing an outlook on solutions and future impact in the field. 
\end{itemize}

The remainder of this paper is structured as follows. Section \ref{sm} presents a high-level system model for EV charging infrastructure. Section \ref{at} formalizes the system's attack surface. Section \ref{det} provides an overview and categorization of the state of the art, with an emphasis on dataset usability, methods, and approaches. Section \ref{oc} discusses open challenges in light of current dataset availability and limitations, and identifies critical future directions to further catalyze innovation in the field. Finally, Section \ref{conc} concludes the paper.  
\section{System Model}
\label{sm}
\begin{figure*}
\centering
    \includegraphics[width=0.7\textwidth]{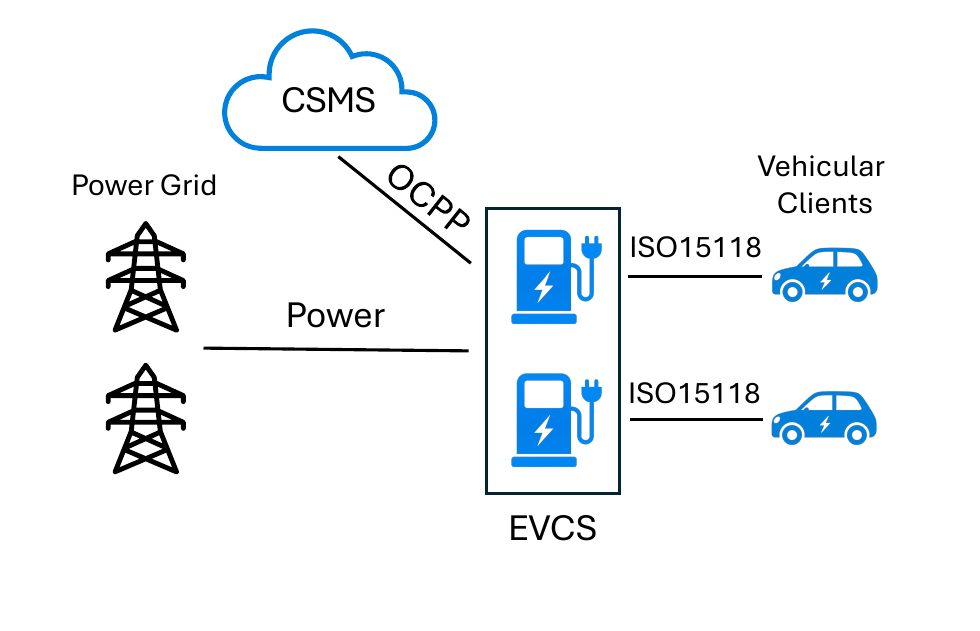}
    \caption{EVSE System Model}
    \label{fig:system-model}
\end{figure*}

A high-level overview of the EV charging ecosystem is presented in Fig. \ref{fig:system-model}. As shown in this figure, EV charging systems comprise various components, including the Electric Vehicle Charging Station (EVCS), the Charging Station Management System (CSMS), the power grid, and vehicular clients. In terms of functionality, the EVCS encompasses both network communication and control, as well as power-delivery hardware known as Electric Vehicle Supply Equipment (EVSE), whereas the CSMS is a cloud management platform for all aspects of user charging sessions, including power flow and pricing. The interaction among all these entities is enabled by various protocols, including the Open Charging Point Protocol (OCPP) and the ISO15118 standard. The OCPP protocol is the foundation of communication between the CSMS and the EVCS and runs over websockets to exchange JSON values throughout the charging session. Inherently, the presence of various subsystems and numerous protocols, coupled with the scalability of clients and infrastructure, can quickly lead to complex network topologies with vast attack surfaces.

\section{EVSE Attack Surface}
\label{at}
Due to its critical positioning between EVs and the power grid, EVSE presents an attack surface with wide-reaching significance. Man-in-the-middle attacks or even compromised charging stations can lead to a host of security threats. As with many IoT devices, vulnerabilities in EVCS are quite common and can have particularly devastating effects \cite{Sayed2025} \cite{Elmo2023}. The risks of these attacks include manipulation of pricing markets, over- or under-billing, or even load shifting attacks against the power grid. Additionally, Denial of Service (DoS) attacks have arisen as a significant issue for EVSE network architectures \cite{Elmo2023}. 

With OCPP being the main application protocol facilitating communication between EVSE and CSMS, security of its implementation has only become more crucial. OCPP has seen numerous security improvements over the last ten years, resulting in an evolving attack surface. The major differences between versions are highlighted in Table \ref{tab:ocpp}. Most significantly, OCPP 2.0.1 adds the requirement of TLS encryption, whereas OCPP 1.6 previously did not enforce it. Even OCPP 2.1 still only requires TLS v1.2 which has known vulnerabilities in many of its possible configurations. Proper configuration and upgrading of OCPP is particularly important due to the level of trust placed in this communication channel between the CSMS and EVSE.

\begin{table*}[!t]
\renewcommand{\arraystretch}{1.3}
\caption{Comparison of OCPP Versions \cite{Open_Charge_Alliance_2026}}
\label{tab:ocpp}
\centering
\footnotesize
\begin{tabular}{|p{2.5cm}|p{3cm}|p{3cm}|p{3cm}|}
\hline
 & \textbf{OCPP 1.6} & \textbf{OCPP 2.0.1} & \textbf{OCPP 2.1} \\
\hline\hline
\textbf{TLS}            & Supported but not enforced   & Requires TLS v1.2      & Requires TLS v1.2            \\ \hline
\textbf{Authentication} & RFID                         & RFID, credit card, PIN & Adds support for authorization cache \\ \hline
\textbf{Message Format} & JSON and SOAP                & JSON via websockets    & JSON via websockets          \\ \hline
\textbf{ISO15118 Support} & No       & Yes                    & Yes                        \\ 
\hline
\end{tabular}
\end{table*}

\section{Detection Strategies}
\label{det}
Given the criticality of systems that could face severe downstream impacts if the EVSE and associated infrastructure were compromised, significant effort has been devoted to developing strategies for detecting cyberthreats and cyberattacks. This survey will focus specifically on intrusion detection and, more broadly, anomaly detection. The majority of works follow two main approaches: network-based intrusion detection or hardware-based anomaly detection. Given the widespread use of ML/AI, it is important to critically assess these works in the context of the data used to develop them. Table \ref{tab:literature} presents an overview of the surveyed works, with an emphasis on the datasets, ML/AI modeling, and the approach used.

\begin{table*}[!t]
\renewcommand{\arraystretch}{1.3}
\caption{Literature Review of Machine Learning Approaches for EV Charging Security}
\label{tab:literature}
\centering
\footnotesize
\begin{tabular}{p{1cm}cp{3cm}p{2.5cm}cc}
\hline
\textbf{Citation} & \textbf{Year} & \textbf{Dataset} & \textbf{Best Model}& \textbf{Network Monitoring} & \textbf{Hardware/Power Monitoring} \\
\hline\hline
\cite{Buedi2024} & 2024 & CICEVSE2024 (dataset paper) & RF, KNN, SVM & \checkmark & \checkmark \\
\hline
\cite{Purohit2024} & 2024 & CICEVSE2024 & DNN & \checkmark &  \\
\hline
\cite{Masum2024} & 2024 & CICEVSE2024 & RF &  & \checkmark \\
\hline
\cite{Thapa2025} & 2025 & CICEVSE2024 & LSTM & \checkmark & \checkmark \\
\hline
\cite{Tanyıldız2025} & 2025 & CICEVSE2024 & GAN &  & \checkmark \\
\hline
\cite{Almadhor2025} & 2025 & CICEVSE2024 & Hybrid LSTM-RNN & \checkmark & \checkmark \\
\hline
\cite{Kumar2025} & 2025 & CICEVSE2024 & XGBoost &  & \checkmark \\
\hline
\cite{Rahman2025} & 2025 & CICEVSE2024 & CNN-LSTM hybrid & \checkmark & \checkmark \\
\hline
\cite{Makhmudov2025} & 2025 & CICEVSE2024 & Adaptive Random Forest & \checkmark &  \\
\hline
\cite{Li2025} & 2025 & CICEVSE2024 & GNN &  & \checkmark \\
\hline
\cite{ElKashlan2023} & 2023 & IoT-23 & Decision Table and Filtered Classifier & \checkmark &  \\
\hline
\cite{Dehrouyeh2024} & 2022 & CICIDS2017 & MLP & \checkmark &  \\
\hline
\cite{Basnet2020} & 2020 & CICIDS 2018 & LSTM & \checkmark &  \\
\hline
\cite{Jahangir2024} & 2024 & ACN-Data & AE/2D CNN & \checkmark &  \\
\hline
\cite{Terruggia2025} & 2025 & Real-world dataset from RSE & LSTM-AE & \checkmark &  \\
\hline
\cite{Sarieddine2024} & 2024 & Real-world dataset from Hydro-Quebec & Convolutional LSTM &  & \checkmark \\
\hline
\cite{Kabir2021} & 2021 & Simulated for study & BPNN & \checkmark &  \\
\hline
\cite{Morosan2017} & 2017 & Simulated for study & BPNN & \checkmark &  \\
\hline
\cite{Jacob2025} & 2025 & Simulated for study & Transformer-based deep learning &  & \checkmark \\
\hline
\cite{Mitikiri2025} & 2025 & Simulated for study & LSTM-AE &  & \checkmark \\
\hline
\cite{Abazari2024} & 2024 & Simulated for study & CNN &  & \checkmark \\
\hline

\end{tabular}
\end{table*}

\subsection{Network Monitoring}

One of the most widely used datasets for EVSE intrusion detection systems is the CICEVSE2024 dataset introduced by Buedi \textit{et al.} \cite{Buedi2024}. This dataset includes DoS and recon attacks in OCPP and ISO15118 traffic, and adds cryptojacking and backdoor attacks to the HPC and kernel events data, as well as to the power consumption data. For this study, works using the CICEV2023 dataset have not been included, as the CICEVSE2024 is more comprehensive and provides a greater variety of attack scenarios. 

Thapa \textit{et al.} use TinyML to deploy an LSTM for classifying attack vectors in CICEVSE2024 \cite{Thapa2025}. Almdhor \textit{et al.} classify attacks in CICEVSE2024 using a variety of deep learning models, including LSTM, RNN, LSTM-RNN, GRU, and transfer learning. The accuracy achieved in this study falls below 90\% for most models \cite{Almadhor2025}. Purohit and Govindarasu present a federated learning anomaly detection system built with a DNN, achieving about 97\% accuracy \cite{Purohit2024}. Rahman \textit{et al.} train several deep learning models to detect attacks, with accuracy also around 97\% \cite{Rahman2025}. Makmudov \textit{et al.} train an online Adaptive Random Forest model for both binary and multi-class attack detection \cite{Makhmudov2025}. 

There are a handful of works on EVSE network monitoring, mainly focused on OCPP, with some using physical charging equipment to generate data \cite{Terruggia2025, Morosan2017}, others using EV charging session datasets \cite{Jahangir2024}, and some simulating power grid attacks through OCPP message manipulation \cite{Kabir2021}. Kabir \textit{et al.} train a BPNN on EVSE actions to detect coordinated switching attacks, in which EVSE power oscillations could destabilize the power grid \cite{Kabir2021}. Morosan and Pop train a BPNN to detect non-compliant or randomly generated OCPP 1.5 (legacy version of the protocol) traffic. Their work does not simulate any attacks but instead aims to detect traffic that appears out of the ordinary \cite{Morosan2017}. Jahangir \textit{et al.} train an autoencoder on charging session start and end times and power demand to detect manipulation of energy prices \cite{Jahangir2024}. Terruggia \textit{et al.} train an LSTM Autoencoder on data derived from typical OCPP traffic to model baseline behavior in the network. The model was tested against DoS attacks and achieved high accuracy \cite{Terruggia2025}.

A subset of works leverage widely used IoT datasets to draw parallels between conventional IoT networks and EVSE environments. Notably, although these studies do not use EVSE-specific datasets, their results demonstrate the robustness and quality of results achievable in similar IoT networks. ElKashlan \textit{et al.} use the IoT-23 dataset to train a Decision Table and a Filtered Classifier to detect DDoS attacks with nearly 100\% accuracy \cite{ElKashlan2023}. Dehrouyeh \textit{et al.} train both Random Forest and MLP models on the CICIDS2017 dataset, achieving nearly 100\% accuracy and demonstrating their utility in a TinyML deployment \cite{Dehrouyeh2024}. Basnet and Ali use CICIDS2018 to train a DNN and an LSTM model to detect multiple classes of attacks in the dataset, with accuracy between 98\% and 100\% depending on the attack \cite{Basnet2020}.

\subsection{Power and Hardware Monitoring}

CICEVSE2024 is also a standard dataset for power and hardware monitoring applications. In addition to packet captures, the EVSE-B section of the CICEVSE2024 dataset includes hardware performance counter (HPC) and kernel events, as well as power data. A distinguishing feature of the HPC and power data is the inclusion of backdoor and cryptojacking attacks, along with the previously reported recon and DoS events. In the original dataset paper, Buedi \textit{et al.} present several traditional ML models for both binary and multiclass classification on HPC and kernel event data. Their initial work achieved 91.3\% accuracy with a KNN model for binary classification and 78.9\% with a Random Forest for multiclass classification \cite{Buedi2024}.

Continuing with the traditional ML approaches, Masum \textit{et al.} achieve 93.4\% accuracy with Random Forest on the HPC and kernel event data, significantly outperforming \cite{Buedi2024} in multiclass classification using the same algorithm \cite{Masum2024}. Similarly, Kumar \textit{et al.} use both the power and host events sections of EVSE-B and perform binary classification focused specifically on DoS attacks. They employ several models, but the best-performing on both sections is XGBoost, achieving 98.6\% accuracy on the host events and 94.5\% on the power measurements. This study attempts true anomaly detection using both autoencoder and isolation forest models, achieving some success with the host event data but struggling with the power data. 

Neural Network solutions performed better overall on the HPC portion of the dataset. Li \textit{et al.} focus solely on the HPC data in EVSE-B, training a GNN for multiclass classification and achieving 97.6\% accuracy \cite{Li2025}. Thapa \textit{et al.} achieve the best performance on the HPC data, with 99.8\% accuracy. They use an LSTM model for multiclass classification \cite{Thapa2025}. The work of Tanyıldız \textit{et al.} analyzes the CICEVSE2024 power data and trains a GRU-GAN model to predict the time until a potential attack, achieving a mean absolute error of 0.0281. While this is a fundamentally different goal, they demonstrate that useful insights can still be gained from this portion of the data \cite{Tanyıldız2025}. 

Aside from the CICEVSE2024 data, there is no standard dataset for power monitoring. Studies generally rely on simulations of charging sessions from third-party sources. Sarieddine \textit{et al.} start with data from Hydro-Québec and introduce attack simulations into their testbed, focusing on the impact on the power grid. By combining power measurement data with EVCS event logs, they generate a dataset of oscillatory load attacks launched by compromised charging stations. They train a Convolutional LSTM on this data to detect attacks with 99.4\% accuracy \cite{Sarieddine2024}. Likewise, Jacob \textit{et al.} train a Transformer-based deep learning model to detect irregular power flow from the grid to EV charging stations. They model and simulate data from Google Maps charging-station activity reports. They report 98.7\% and 94.4\% accuracy in their two test networks \cite{Jacob2025}. Abazari \textit{et al.} also detect grid-targeting attacks, specifically in wind power grids. They train a CNN on simulated voltage data, achieving 98\% accuracy in detecting benign, malicious, and faulted data \cite{Abazari2024}. Mitikiri \textit{et al.} approach the problem from the vehicle side by monitoring the charging port. They train an LSTM-based AutoEncoder to detect anomalies in current draw from the charging port, achieving 98.5\% \cite{Mitikiri2025}.

\subsection{Key Takeaways}
Based on the aforementioned works, it is clear that CICEVSE2024 has become the standard dataset for both network and power monitoring applications since its introduction. This dataset was the first to provide comprehensive coverage of a suite of attack vectors and the associated power and hardware-related information. This inclusion of diverse data aspects was a critical step in advancing the state of the art, enabling more accurate modeling of the intricacies of real-world EVSE infrastructure than previous datasets. It was also a major step toward the democratization of data in the field, as this reference dataset served as a foundation for many works and ML-based applications. This is evidenced by the variety of models that were built, ranging from classical ML to deep learning, recurrent neural networks, and even transformers. However, despite its usability, there are noticeable limitations in terms of scalability, variability, and modeling capabilities that must be addressed to continue advancing this ever-evolving field.

\section{Open Challenges and the Future of EVSE Security}
\label{oc}
\begin{figure*}[!t]
    \centering
    \includegraphics[width=0.8\textwidth]{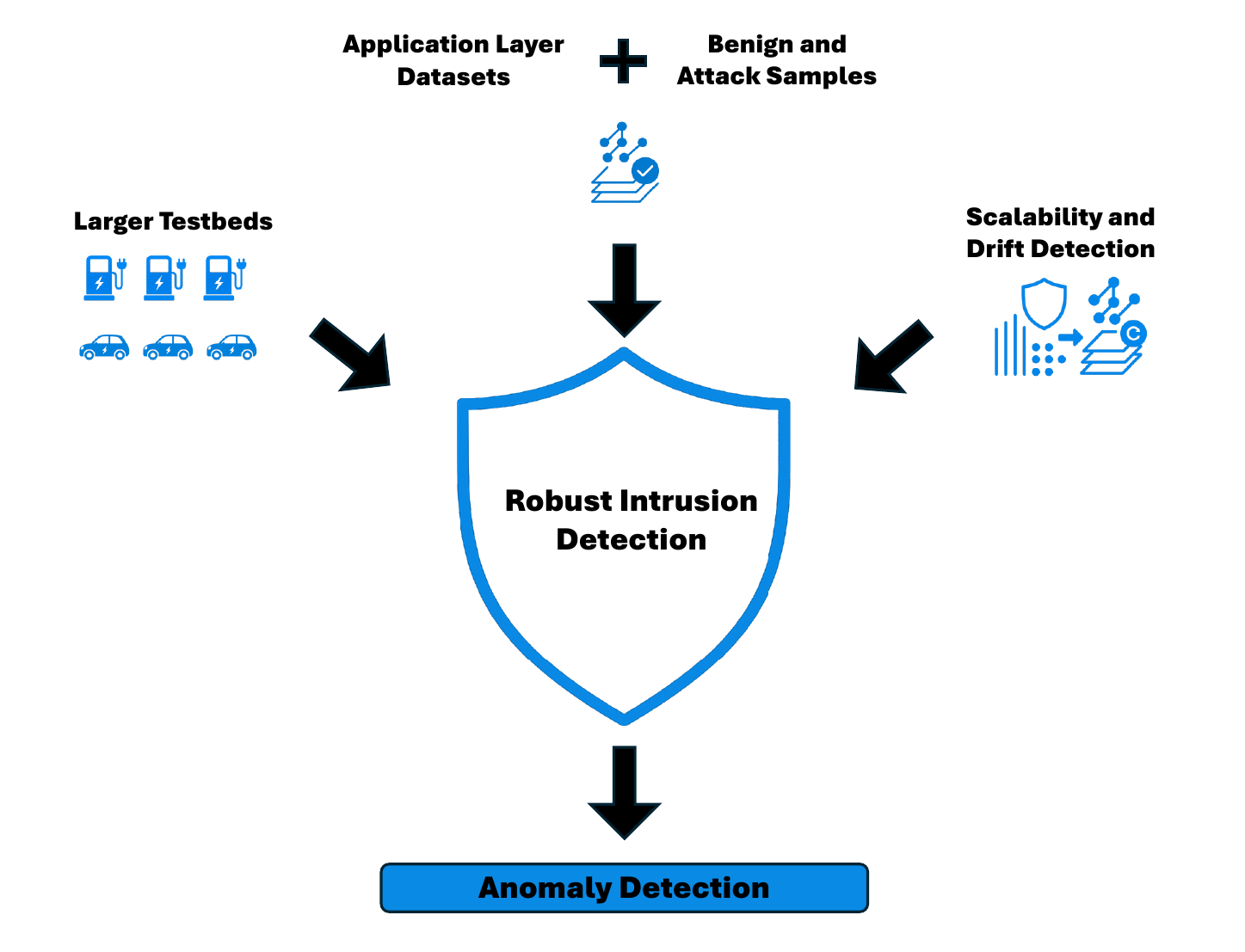}
    \caption{Future of EVSE Security}
    \label{fig:figure2}
\end{figure*}

This section provides an analysis of the current limitations in this field of research, as well as a summary of the solutions required to fill these gaps. As a whole, addressing these gaps will contribute to the final goal of implementable anomaly detection with visibility at the application layer. Figure \ref{fig:figure2} provides a visualization of the components addressed in this section.

\subsection{Dataset Limitations}
Cyber intrusion detection models are only as good as the data on which they are trained. As novel attacks emerge, thorough datasets are needed to document and model these evolutions. There do exist many comprehensive IoT and IDS-oriented datasets, but very few of them are specific to the EVSE domain. CICEVSE 2023 is one such dataset, although it only models DDoS attacks \cite{Kim2023}.

The CICEVSE2024 dataset is comprehensive as it pertains to attacks visible at the network layer. However, in the context of developing deployment-ready anomaly detection models, it has some notable shortcomings. First, the testbeds are very small, each only including one charger and minimal network architecture. While there is variety in the implementation between EVSE-A and EVSE-B, the scale of the network limits the amount of benign data that is able to be represented by the network flows. Since network flows model complete connections, normal charging sessions with no added attack traffic are easily represented in very few data samples. Furthermore, the EVSE-B packet captures do not even contain any benign data. Real-world EVSE deployments are likely to contain many more devices, therefore, generating more traffic and exhibiting more susceptibility to stealthy attack scenarios.

Secondly, the CICEVSE2024 network communications are properly secured in transit, meaning that no application data is exposed in the packet captures. This is not a problem for detecting the DoS and reconnaissance attacks simulated in the data, but it does exclude the possibility of any IDS model learning baseline behavior at the application layer. While a fundamentally different type of dataset is needed for this task, the impacts of application protocol manipulation attacks are too great to ignore. Odimegwu \textit{et al.} have already begun to address this gap with their OCPP dataset \cite{Odimegwu2025}. As these complex and often stealthier data injection and manipulation attacks mature, it is crucial that they are documented and studied by cyber defenders.

EVSE security would greatly benefit from the addition of new public datasets that capture these missing aspects of existing data. These future datasets should begin by including a wider range of attacks that may be encountered in EVSE systems. As many of these attacks are invisible at the network layer, application layer data is a priority. Since these attacks may occur as a result of EVCS component compromise, not just network infiltration, IDS models should be able to recognize anomalies originating from trusted components as well as from separate malicious entities in the network.

\subsection{Anomaly Detection and Behavior Modeling}
The emergence of new attack vectors is a constant issue in cybersecurity, and one that is inadequately addressed by standard ML/AI attack classification models. As such, more attention should be given to models that generalize and alert on behavior rather than on specific attack signatures. In the field of EVSE security, this approach is largely missing. Besides the few studies that attempt true anomaly detection such as \cite{Jahangir2024} and \cite{Terruggia2025}, attack detection has largely been limited to binary or multiclass classification. There is a need for systems modeling the normal behavior of the various EVSE protocols. These systems should then be tested against attacks designed blend in with normal protocol behavior.

Another key consideration in behavior modeling is the scalability and robustness of attack detection models. As EVSE network topology changes and protocols are updated, intrusion detection systems should still be able identify attacks by modeling baseline operation. Again, accomplishing this goal will require larger testbeds and more protocol-specific insights. Without either aspect, intrusion detection is severely limited in scope and practical applicability.

\section{Conclusion}
\label{conc}

The work presented in this paper summarizes the evolving landscape of EVSE cybersecurity, with a focus on ML/AI-based intrusion and anomaly detection. While many datasets exist, both specific to the EVSE environment and more broadly applicable to IoT systems, the CICEVSE2024 dataset has emerged as a standard for network, power, and hardware monitoring tasks. This dataset has enabled numerous studies using a variety of modeling techniques, ranging from fundamental ML models (\textit{e.g.,} decision trees, kNN, \textit{etc.}) to more complex neural network architectures (\textit{e.g.,} RNNs and Transformers). However, several considerations must be addressed as this field continues to progress. Specifically, in the context of datasets, challenges arise regarding the scale of testbeds used to generate data and the depth of the attack vectors they contain. In terms of methodology, anomaly detection methods that model expected system behavior and are robust to unseen attack vectors are a key path toward securing the future of the EVSE ecosystem. By addressing all aspects of the ML/AI pipeline, future cybersecurity defenses can be better prepared to handle novel attack patterns employed by adversaries.

\bibliographystyle{IEEEtran}
\bibliography{sample}

\end{document}